\documentclass[final,5p,times,twocolumn]{elsarticle}

\usepackage{amssymb}
\usepackage{makecell}
\usepackage{booktabs}
\usepackage{makecell}
\usepackage{threeparttable}
\usepackage{graphicx}
\usepackage{epstopdf}
\usepackage{float}
\usepackage{amsmath,amsfonts}
\usepackage{algorithm}
\usepackage{algpseudocode}
\usepackage{array}
\usepackage[caption=false,font=normalsize,labelfont=sf,textfont=sf]{subfig}
\usepackage{textcomp}
\usepackage{stfloats}
\usepackage{verbatim}
\usepackage{color}
\usepackage{multirow}
\usepackage[hyphens]{url}
\usepackage{breakurl}
\usepackage{amssymb}
\usepackage{verbatim}

\usepackage{bm} 
\usepackage{color}
\newcommand{\xx}[1] {\textbf{{\color{red}{#1}}}}
\newcommand{\xxcomment}[1] {{\color{blue}{#1}}}

\journal{Applied Acoustics}

\begin{document}

\begin{frontmatter}

\title{Speaker Recognition Using Isomorphic Graph Attention Network Based Pooling\\
on Self-Supervised Representation\tnoteref{funding}}
\tnotetext[funding]{This work was supported by the National Natural Science Foundation of China under Grants 62071242, the China Postdoctoral Science Foundation under Grant 2022M711693, and Postgraduate Research and Practice Innovation Program of Jiangsu Province (KYCX23\_1034).}

\author[author1]{Zirui Ge\corref{cor1}}
\address[author1]{School of Communication and Information Engineering, Nanjing University of Posts and Telecommunications, Nanjing 2100023, Jiangsu, China}
 \ead{2022010211@njupt.edu.cn}
\author[author2]{Xinzhou Xu}
\address[author2]{School of Internet of Things, Nanjing University of Posts and Telecommunications, Nanjing 2100023, Jiangsu, China}
\author[author1]{Haiyan Guo}

\author[author1]{Tingting Wang}

\author[author1]{Zhen Yang\corref{cor1}}
\ead{yangz@njupt.edu.cn}

\cortext[cor1]{Corresponding author(s).}

\begin{abstract}
The emergence of self-supervised representation (i.e., wav2vec 2.0) allows speaker-recognition approaches to process spoken signals through foundation models built on speech data. Nevertheless, effective fusion on the representation requires further investigating, due to the inclusion of fixed or sub-optimal temporal pooling strategies. Despite of improved strategies considering graph learning and graph attention factors, non-injective aggregation still exists in the approaches, which may influence the performance for speaker recognition. In this regard, we propose a speaker recognition approach using Isomorphic Graph ATtention network (IsoGAT) on self-supervised representation. The proposed approach contains three modules of representation learning, graph attention, and aggregation, jointly considering learning on the self-supervised representation and the IsoGAT. Then, we perform experiments for speaker recognition tasks on VoxCeleb1\&2 datasets, with the corresponding experimental results demonstrating the recognition performance for the proposed approach, compared with existing pooling approaches on the self-supervised representation.


\end{abstract}



\end{frontmatter}

\begin{figure*}[t]
\centering
\includegraphics[width=7.2in]{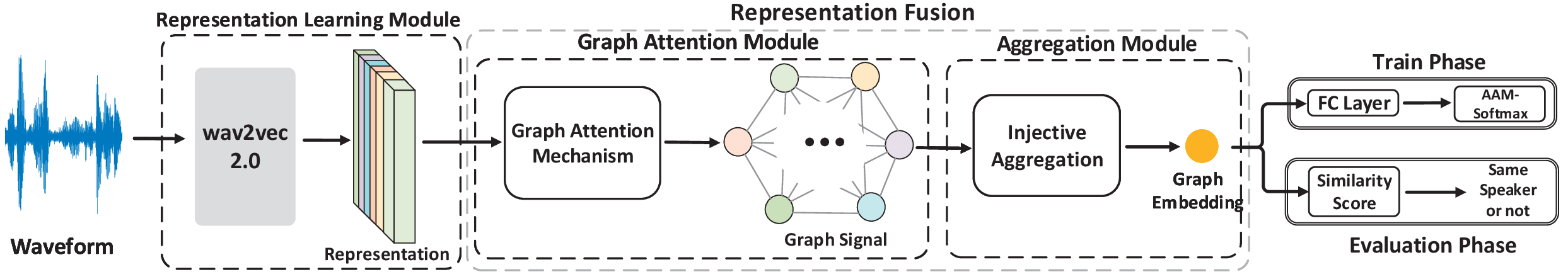}
\caption{A diagrammatic overview of the proposed IsoGAT approach, including three modules of representation learning, graph attention, and aggregation, based on the self-supervised representation. }
\label{framework}
\end{figure*}

\section{Introduction}
\label{1}
As a critical task in biometric authentication, \emph{Speaker Recognition} (SR) requires the process of verifying the identity of a speaker, through their voiceprints in speech~\cite{Hansen15-SRB,Lin22-MRL}, owing to its uniqueness and accessibility~\cite{Ref45}. 
Following previous SR research with shallow structures, the advent of \emph{Deep Neural Networks} (DNNs) has resulted in the requirement of large-size labeled training data for deep SR models~\cite{Nagrani20-VLS}. 
Nevertheless, it is usually insufficient to acquire knowledge only from intra-domain labeled samples in the perspective of cognitive learning~\cite{Aakur19-APP}, and hence may call for the inclusion of information transfer from inter-domain self-learning process as upstream tasks~\cite{Ref46, Mehrish23-ARO}. \xx{In light of this requirement,} the emerged \emph{Self-Supervised Learning} (SSL) provides the possibility to build pre-trained models on unlabeled data for downstream tasks~\cite{Liu21-TSS,Baevski22-DAG,Zaiem22-PTS,Achyut21-SSL}, leading to the emergence of wav2vec 2.0 framework within the scope of spoken signal processing~\cite{Ref3}.

Further research related to the wav2vec 2.0 framework addresses downstream speech-analysis tasks, using the SSL-based model pre-trained on large corpora~\cite{Wang23-DSI,Latif-SSA, Hsu21-HSS}, which has been proved effective for these tasks. 
\xx{For SR-related tasks, \cite{Ref4} regards wav2vec 2.0 as an audio encoder to extract speaker and language information for SR and language identification. Then, \cite{Chen2022-LSS} employs the ECAPA-TDNN~\cite{Desplanques20-EEC} as a downstream module to process the representations learned from the pre-trained models, while \cite{Ref6} investigates the effectiveness of different classical pooling strategies for wav2vec 2.0's outputs.}
Further, in paralinguistic cases, \emph{Pepino} et al. considers the pre-trained wav2vec 2.0 models for \emph{Speech Emotion Recognition} (SER), through weighting the outputs of the multiple layers from the models~\cite{Ref5}.

Along with the wav2vec 2.0 framework, pooling operations on sequential signals require further design for adaptive-fusion strategies on speech sequences, despite of the previous research on classical pooling~\cite{Ref6,Ref5}. 
\xx{Within these pooling strategies, graph-learning based approaches have been investigated for adapting to speech analysis. }
The works of \cite{Ref8,Ref9,Ref10} have shown that a speech utterance can be reformulated as graphs processed with \emph{Graph Signal Processing} (GSP) theory~\cite{Ortega18-GSP} in irregular spaces. As a nonlinear extension of GSP, \emph{Graph Neural Network} (GNN) is initially utilized as a backend feature-fusion method on \emph{Residual Network} (ResNet) and RawNet2 models for SR~\cite{Ref19, Ref20}. 
In detail, \emph{Jung} et al. considers segment-wise speaker embeddings as the input to \emph{Graph ATtention networks} (GATs)~\cite{Ref19,graphGAT}, and then \emph{Shim} et al. design directed graphs using GATs and obtain the final graph representation using U-Net architectures~\cite{Ref20}. 
In addition, \cite{Ref21} employs the same GAT architecture in \cite{Ref19} to implement speaker anti-spoofing.

Nevertheless, existing works include two deficiencies for addressing SR tasks using self-supervised representation. 
First, existing pooling strategies on self-supervised representation for SR tasks mainly rely on fixed or sequential temporal fusion, which may fail to effectively describe the complex intrinsic relationship between these low-level components. 
Second, existing graph learning based approaches for speech processing contain non-injective aggregation, possibly hindering the performance of GNNs. 
To this end, we propose a speaker recognition approach using \emph{Isomorphic Graph ATtention network} (IsoGAT) on self-supervised representation, which targets at these deficiencies through considering isomorphism in GAT for SR tasks.

Within the proposed approach, we first perform representation learning to present a speakers' identity-aware transformed space, in order to obtain  low-level self-supervised representations, instead of original latent embeddings. 
Further, the obtained representations are fed to a graph attention module considering GAT with cosine similarity (noted as ‘GATcosine’)~\cite{Ref29}, to model the low-level components' relationship. 
Afterwards, we design an aggregation module to perform adaptive pooling through building an improved GAT model, with injective aggregation and updating function for the GAT layer, based on \emph{Graph Isomorphism Network} (GIN)~\cite{GIN}.

Further, we make comparison between the proposed and highly-related existing works. 
Compared with classical pooling methods~\cite{Ref4,Ref5,Ref6}, GAT-based approaches model the representations as graph
signals, and assign learnable weights between low-level components in an utterance, which outperforms \emph{Recurrent Neural Networks} (RNNs) relying on temporal correlation~\cite{Ref44}. 
Compared with other GAT-based approaches~\cite{ Ref20}, we follow the GATcosine similarity between vertices and design an isomorphic aggregation strategy.  
We also observe a related work~\cite{Gharaee21-GRL} proposing a direct fusion of GAT and GIN, while our approach  utilizes GIN to address the non-injective aggregation issue specific to GATcosine.

The main contributions of this paper are presented as follows:

\begin{itemize}
\item[$\bullet$] We propose a speaker recognition approach using IsoGAT-based pooling on self-supervised representation, containing the modules of representation learning, graph attention, and aggregation. 

\item[$\bullet$] In the graph attention module, we propose to include a GAT with cosine similarity, for the purpose of learning low-level attention weights.   

\item[$\bullet$] In the aggregation module, we propose to set GIN-based injective functions in the GAT layers, in order to enhance the system's expression ability. 

\end{itemize}

The remainder of this paper is organized as follows. Section~\ref{sec-rw} introduces the related works, while the proposed approach is detailed in Section~\ref{sec-met}. Then, Section~\ref{sec-es} and ~\ref{sec-er} presents the experimental setups and results, respectively. Finally, Section ~\ref{sec-con} concludes the paper.

\begin{figure*}[t]
\centering
\includegraphics[width=5.8in]{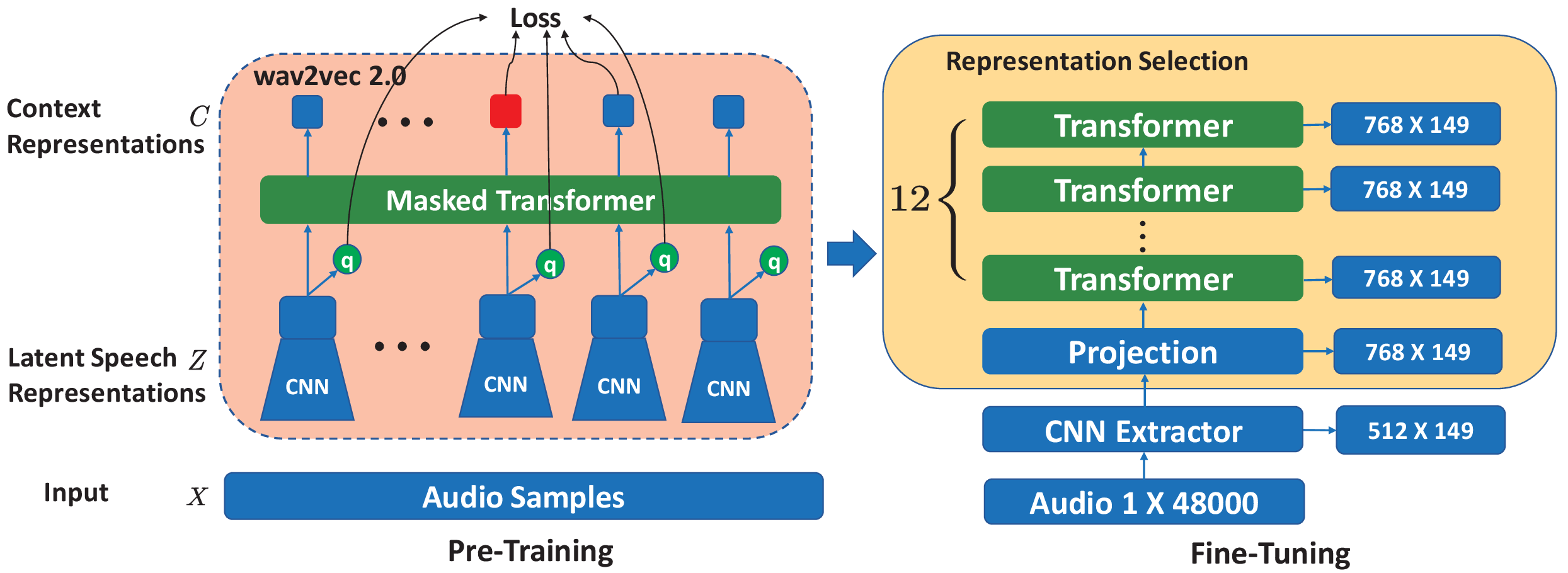}
\caption{A diagrammatic overview of the pre-training and fine-tuning procedures in the representation learning module, with the left part indicating the pre-training phase and the right part corresponding to the fine-tuning phase. }
 \label{wav2vec2}
\end{figure*}

\section{Related Works}
\label{sec-rw}

\subsection{Self-Supervised Speech Representation Learning.}

The emerging \emph{Self-Supervised Learning} (SSL) leverages inherent data characteristics as the labels to learn effective representation for downstream tasks~\cite{Liu21-TSS,Baevski22-DAG}. In speech processing, conventional SSL methods comprise two stages to learn speech signals' representations: In the first stage, SSL is leveraged to pre-train a representation model, typically referred as upstream tasks or foundation models~\cite{Bommasani21-OTO,Ref46}. Then, downstream tasks either directly employ the learned representation from the fixed model, or perform fine-tuning using the pre-trained models for the tasks in target domains~\cite{Refssl}.

SSL for speech representation learning can be categorized into discriminative and generative methods. Discriminative methods directly learn representations mainly using metric learning-based objectives, while generative ones aim to learn representations through reconstructing the input speech data or predicting masked parts of the data~\cite{Ref46,Chung21-SAO}, typically including Mockingjay~\cite{RefMockingjay}, Audio ALBERT \cite{RefALBERT}, and \emph{Transformer Encoder Representations from Alteration} (TERA)~\cite{RefTERA}. As for discriminative models, considering the inclusion of \emph{Contrastive Predictive Coding} (CPC)~\cite{Aaron19-RLW}, previous works present the SSL models of wav2vec~\cite{schneider2019wav2vec}, wav2vec-C~\cite{Sadhu21-WAS} and VQ-wav2vec~\cite{vq-wav2vec}.

\subsection{Graph Signals and Graph Neural Network.}

The existing research on GSP extends classical \emph{Digital Signal Processing} (DSP) technologies to irregular-structure graph signals~\cite{Ref25, Ref26}. 
GNN initially emerges as nonlinear extensions of graph filters, achieved through introducing pointwise nonlinearity to the processing pipeline~\cite{ruiz2021graph}. 
Various GNN-based frameworks have been proposed, including \emph{Message Passing Neural Network} (MPNN)~\cite{Ref17}, GAT~\cite{graphGAT,Ref29}, and GIN~\cite{GIN}. 
Improved GNNs in spatial domains mainly follow the MPNN framework firstly proposed for chemical prediction~\cite{Ref17}, where the vertex representations are iteratively updated by means of aggregating the representations of each vertex's neighbors~\cite{Nikolentzos-PMS}. 
As a common framework, MPNN unifies many existing methods of \emph{Gated Graphed Sequence Neural Networks} (GGS-NNs) \cite{ruiz2020gated}, and \emph{Deep Tensor Neural Networks} (DTNNs)~\cite{NIPS2015_f9be311e}.

GAT also follows the MPNN framework which incorporates the self-attention mechanism~\cite{Ref18} into the propagation step, and obtains new vertex features via weighting the features of the corresponding neighboring vertices using attention coefficients. 
The graph attention mechanism in GAT contains two strategies: The first one leverages explicit attention mechanism to obtain the attention weights, for instance, the cosine similarity between different vertex pairs~\cite{Ref29}. 
In contrast, the other strategy does not rely on any prior information, and leverages complete parameters learning to gain attentive weights~\cite{graphGAT}. Further, \cite{GIN} proposes a GIN framework through using summation and a \emph{Multi-Layer Perceptron} (MLP), to ensure the GNN possess as large discriminative power as in the \emph{Weisfeiler-Lehman} (WL) test~\cite{WL}.



\section{METHODOLOGY}
\label{sec-met}

The proposed IsoGAT contains three modules of representation learning, graph attention, and aggregation, as presented in Figure~\ref{framework}. 
Within the proposed approach, we first employ the self-supervised pre-training model (i.e., wav2vec 2.0) to obtain low-level representation for further fine-tuning. 
Then, the representation is input to the GATcosine-based graph attention module in order to generate weighted representation. 
Finally, the aggregation module aims to perform fusion on the low-level representation using isomorphic message passing.


\subsection{Representation Learning Module}
\label{subsec-RLM}

We consider the self-supervised pre-trained wav2vec 2.0 model as the representation learning module~\cite{Ref3,Ref6}. 
The main body of the model consists a CNN based feature encoder, a Transformer-based context network, a quantization sub-module, and a contrastive loss. 
The feature encoder includes $7$ blocks with the respective kernel sizes of $(10, 3, 3, 3, 3, 2, 2)$ and 
the respective strides of $(5, 2, 2, 2, 2, 2, 2)$, followed by a layer normalization and a \emph{Gaussian Error Linear Unit} (GELU) activation function~\cite{gelu}.

As shown in the pre-training stage in Figure~\ref{wav2vec2}, the convolutional layers output latent speech representations, using the input raw speech. 
Then, we perform projection on the representations into a new space, before proceeding next steps. 
The following context network contains $12$ Transformer blocks, in which we first add the relative positional embeddings to the masked speech representations. Then, the Transformer blocks contextualize the masked representations and finally generates context representations. Through jointly considering the context representations and the projected latent speech representations, a combined loss can be formulated using the weighted sum of contrastive and diversity losses. 

The right part of Figure~\ref{wav2vec2} shows the fine-tuning stage, removing the output layer with the loss function in the pre-training stage, add a representation-selection block to process the $13$ output hidden representations (output from the $12$ Transformer layers and the projection). 
As the representation selection may influence its downstream tasks~\cite{Ref5}, we design two strategies for the selection. For an arbitrary utterance sample $\boldsymbol{x}$, the first strategy employs the $F$-dimensional column vector $\boldsymbol{r}_{i,13}$ representing the output of the last Transformer block, for the time step $i=1,2,...,N$ in $\boldsymbol{x}$. The second one aims to weight all the $13$ output representations $\boldsymbol{r}_{i,l_0}$ $(l_0=1,2,\ldots,13)$~\cite{Ref5}, with the corresponding linear trainable weights $\boldsymbol{d}_i=[d_{i,1},d_{i,2},\ldots,d_{i,13}]^T$, leading to the step-$i$ output for the utterance $\boldsymbol{x}$ represented as 
\begin{equation}\label{eq1}
\tilde{\boldsymbol{x}}_i=(\boldsymbol{d}_i^T\boldsymbol{e}_{13})^{-1}[\boldsymbol{r}_{i,1},\boldsymbol{r}_{i,2},\ldots,\boldsymbol{r}_{i,13}]\boldsymbol{d}_i,
\end{equation}
where $\boldsymbol{e}_{13}$ indicates a $13$-dimensional column vector with all its elements equal to $1$.




\subsection{Graph Attention Module} \label{addlinear}

Let $\mathcal{G}=(\mathcal{V},\mathcal{E})$ be a graph and $\boldsymbol{X}=[\tilde{\boldsymbol{x}}_1,\tilde{\boldsymbol{x}}_2,\ldots,\tilde{\boldsymbol{x}}_N]\in \mathbb{R}^{F\times N}$ be the input vertex representations with $F$ dimensions, where $\mathcal{V}$ represents the set of $N$ vertices, corresponding to the vertex representations. 
$\mathcal{E} =\left\{ \varepsilon_{i,j} \right\} _{i,j\in \mathcal{V}}$ is the set of edges between vertices, such that $\varepsilon_{i,j}=1$ if there is a link between the nodes $i$ and $j$, otherwise $\varepsilon_{i,j}=0$. 
Note that $\boldsymbol{A}$ is the $N\times N$ weighted adjacency matrix of $\mathcal{G}$ and the entries of $\boldsymbol{A}$ correspond to the edge weights $a_{i,j}$ with $i, j=1,...,N$. 
Then, we formulate a graph signal using the output representations of the wav2vec 2.0 model, considering a complete graph with $\varepsilon_{i,j}=1$ for each edge. 


To make the upstream representations adapt to their downstream SR task, we map the obtained embeddings into a space dominated by speaker information via linear transformation, written as 
\begin{equation}\label{eq5}
\boldsymbol{h}^{(0)}_i=\boldsymbol{W}\tilde{\boldsymbol{x}}_i + \boldsymbol{o}, 
\end{equation}
where $\boldsymbol{W}\in \mathbb{R} ^{F'^{(0)}\times F}$ is the projection matrix, and the $F'^{(0)}$-dimensional $\boldsymbol{o}$ is the offset. Then, we obtain the elements of the adjacency matrix $\boldsymbol{A}$ represented as 
\begin{equation}
a_{i,j}=\frac{e^{ \beta \cos \left( \boldsymbol{h}_{i}^{(0)},\boldsymbol{h}_{j}^{(0)} \right) }}{\sum\nolimits_{l\in \mathcal{N}\left(i\right)}{e^{ \beta \cos \left( \boldsymbol{h}_{i}^{(0)},\boldsymbol{h}_{l}^{(0)} \right) }}}, \label{eq6}
\end{equation}
with $i,j=1,2,\ldots,N$, and $\beta $ is a learnable parameter, $\mathcal{N}\left(i\right)$ is the neighbour vertex set of $i$, including vertex $i$. In the GAT layer, the vertex feature is first projected into $F'^{(0)}$ dimensional space via multiplying $\boldsymbol{W}$. Then the attention score is obtained by Equation~\eqref{eq6}.

\begin{figure}[t]
\centering
\includegraphics[width=2.2in]{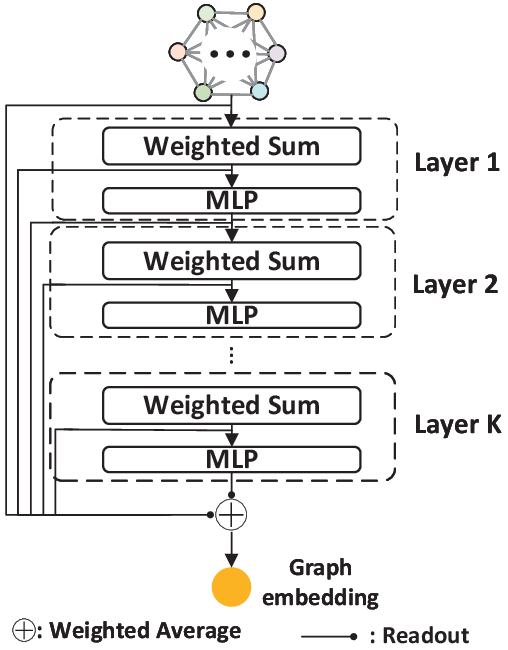}
\caption{An overview of the aggregation module including $K$ layers consisting weighted sum on the vertices' states and an MLP for each layer. }
 \label{aggregation_module}
\end{figure}

\subsection{Aggregation Module}

Afterwards, as shown in Figure~\ref{aggregation_module}, the aggregation module is presented with injective aggregation, considering GATcosine-based aggregation function, with the $k$th-layer hidden state of $i$th vertex with $F'^{(k)}$ nodes, written as
\begin{equation}
\boldsymbol{h}_{i}^{(k)}=\varphi_i(\mathcal{H}^{(k-1)})=a_{i,i}\boldsymbol{h}_{i}^{(k-1)}+\sum\nolimits_{j \in \mathcal{N}\left(i \right), j\ne i}{a_{i,j}\boldsymbol{h}_{j}^{(k-1)}}
. \label{eq7}
\end{equation}
Using the mapping $\varphi_i(\cdot)$, and $\mathcal{H}^{(k-1)}=\{\boldsymbol{h}_{1}^{(k-1)},\boldsymbol{h}_{2}^{(k-1)},\ldots,\boldsymbol{h}_{N}^{(k-1)}\}$ refers to the hidden-state set of the $N$ vertices for the $(k-1)$th layer, where $k=1,2,\ldots,K$ with maximum $K$ layers. Nevertheless, the aggregation scheme in Equation (\ref{eq7}) is non-injective and can be proved as follows.

\par \noindent\textbf{Theorem 1:} 
The aggregation scheme in Equation (\ref{eq7}) is non-injective, \emph{i.e.,} the aggregation scheme can map two vertices with different neighbor representation sets into the same representation. 

\par \noindent \textbf{Proof:} First, we set two inputs of $\check{\mathcal{H}}=\left\{ \check{\boldsymbol{h}}_1,\check{\boldsymbol{h}}_2,\cdots ,\check{\boldsymbol{h}}_N\right\}\subset\mathbb{R}^{F'}$ and $\hat{\mathcal{H}}=\left\{ \hat{\boldsymbol{h}}_1,\hat{\boldsymbol{h}}_2,\cdots ,\hat{\boldsymbol{h}}_N \right\}\subset\mathbb{R}^{F'}$ as the representation set of  for $\mathcal{H}^{(k-1)}$, where $\check{\mathcal{H}}\ne\hat{\mathcal{H}}$. Hence, we aim to examine whether it exists $\check{\mathcal{H}}$ and $\hat{\mathcal{H}}$, such that  $\varphi_i(\check{\mathcal{H}}) = \varphi_i(\hat{\mathcal{H}})$.

Without loss of generality, we set $i=1$ and $\Dot{\boldsymbol{h}}_1=\varphi_1(\hat{\mathcal{H}})$ for the first layer of the hidden states, where $\Dot{h}_{1,1} \ne \Dot{h}_{1,2}$ and $\Dot{h}_{1,1} \ne 0$, $\Dot{h}_{1,2} \ne 0$. We then elaborate an $\check{\mathcal{H}}$ through letting its  
\begin{equation}
\left\{ \begin{array}{l}
	\check{\boldsymbol{h}}_1=\frac{\Dot{h}_{1,1}}{e^{\beta}}{(N-1+e^{\beta})}\boldsymbol{u}_1,\\
	\check{\boldsymbol{h}}_j=\Dot{h}_{1,j}{(N-1+e^{\beta})}\boldsymbol{u}_j\ (j=2,3,\ldots,N), \\
\end{array}\right.
\label{eq1e}
\end{equation}
be one-hot-like vectors, where $\boldsymbol{u}_1$ and $\boldsymbol{u}_j$ refer to the first and the $j$th columns respectively of an $N$-dimensional identity matrix, while $\Dot{h}_{1,1}$ and $\Dot{h}_{1,j}$ are the first and the $j$th element respectively of $\Dot{\boldsymbol{h}}_1$. 
This results in the corresponding elements of the adjacency matrix $\boldsymbol{\check{A}}$ represented as 
\begin{equation}
\left\{ \begin{array}{l}
	\check{a}_{1,1}=\frac{e^{\beta}}{N-1+e^{\beta}}, \\
	\check{a}_{1,j}=\frac{1}{N-1+e^{\beta}}\  (j=2,3,\ldots,N), \\
\end{array}\right.
\label{eq2e}
\end{equation}
through using Equation~\eqref{eq6}, leading to $\varphi_1(\check{\mathcal{H}}) = \Dot{\boldsymbol{h}}_1$. 

Then, we elaborate an $\hat{\mathcal{H}}$ ($\hat{\mathcal{H}}\ne \check{\mathcal{H}}$) through letting its 
\begin{equation}
\left\{ \begin{array}{l}
	\hat{\boldsymbol{h}}_1=\frac{\Dot{h}_{1,2}}{e^{\beta}}{(N-1+e^{\beta})}\boldsymbol{u}_2, \\
	\hat{\boldsymbol{h}}_2=\Dot{h}_{1,1}{(N-1+e^{\beta})}\boldsymbol{u}_1, \\
    \hat{\boldsymbol{h}}_j=\Dot{h}_{1,j}{(N-1+e^{\beta})}\boldsymbol{u}_j\ (j=3,4,\ldots,N), \\
\end{array}\right.
\label{eq1e}
\end{equation}
resulting in the same corresponding elements with $\boldsymbol{\check{A}}$, such that $\varphi_1(\hat{\mathcal{H}})=\Dot{\boldsymbol{h}}_1$. Hence, we obtain $\varphi_1(\check{\mathcal{H}}) =\varphi_1(\hat{\mathcal{H}})= \Dot{\boldsymbol{h}}_1$ for $\check{\mathcal{H}}\ne \hat{\mathcal{H}}$, and consequently $\varphi \left(\cdot \right) $ is non-injective. 
$\blacksquare$

Further, in order to strengthen the expressive ability of GAT, Equation (\ref{eq7}) can be changed into an injective aggregation form with weighted sum on the vertices' states $\boldsymbol{h}_{i}^{(k)}=f^{(k)}(\boldsymbol{m}_{i}^{(k)})$, represented as  
\begin{equation}
\boldsymbol{h}_{i}^{(k)}=f^{(k)}\left( \left( 1+\epsilon ^{(k)} \right) a_{i,i}\boldsymbol{h}_{i}^{\left( k-1 \right)}+\sum\nolimits_{j \in \mathcal{N}\left(i \right),j\ne i}{a_{i,j}\boldsymbol{h}_{j}^{(k-1)}} \right) , \label{GINAGG}
\end{equation}
as a  weighted version of \cite{GIN}. $f^{(k)}(\cdot)$ indicates the $k$th \emph{Multi-Layer Perceptron} (MLP) contained in the improved aggregation form on the weighted sum $\boldsymbol{m}^{(k)}_i$, and $\epsilon^{(k)}$ has many choices, including all irrational numbers. 
Then, we prove the injective property of Equation~\eqref{GINAGG} represented as Theorem 2, with the help of the corollary in \cite{GIN}.

\begin{table}[t]\small
\centering
\begin{threeparttable}
\caption{The description of the validation and test sets employed in the experiments.}
\label{dataset}
\begin{tabular}{lcccc}
\toprule
& \textbf{Validation} & \textbf{Vox1-o} & \textbf{Vox1-e} & \textbf{Vox1-h} \\ \midrule
\# Speakers & $118$    & $40$ & $1\,251$ & $1\,190$ \\
\# Trials & $36$K & $37$K	& $580$K	& $550$K \\
\bottomrule
\end{tabular}
\begin{tablenotes}
			\footnotesize
            \item 1: The numbers of the speakers in Vox1-h are equal on nationality and gender.
			
	\end{tablenotes}
\end{threeparttable}
\end{table}

\begin{table*}[t]\small
\centering
\begin{threeparttable}
\caption{\xx{The EER ($\%$) results of different pooling-based methods employing the self-supervised pre-training representation of wav2vec 2.0, on the three test sets when considering the last-layer and all-layer cases.}} \label{results}
\label{tab1}
\begin{tabular}{cc |c c c c c c}
\toprule
\makecell[l]{\multirow{2}{*}{\textbf{Pooling Types}}} &\makecell[l]{\multirow{2}{*}{\textbf{Methods}} } & \multicolumn{2}{c}{\textbf{Vox1-o}} & \multicolumn{2}{c}{\textbf{Vox1-e}}&\multicolumn{2}{c}{\textbf{Vox1-h}}\\
\multirow{1}{*}{} &{}  & {Last Layer} & {All Layers} &  {Last Layer} & {All Layers} & {Last Layer} & {All Layers} \\
\midrule

 & 
\makecell[l]{\multirow{1}{*}{\emph{Mean} \cite{Ref4, Ref6}}} & $1.75$& $1.96$	& $1.73$ & $2.04$ & $3.30$ & $3.77$ \\ &
\makecell[l]{\multirow{1}{*} {\emph{Maximum} \cite{Ref6}}} & $1.93$ & $1.91$	& $1.91$	& $2.04$ & $3.53$ & $3.53$ \\& 
\makecell[l]{\multirow{1}{*} {\emph{Random}~\cite{Ref6}}} & $1.76$ & $1.95$ & $1.78$ & $1.97$ & $3.37$ & $3.64$ \\
\multicolumn{1}{l}{Classical Pooling} & 
\makecell[l]{\multirow{1}{*} {\emph{First} \cite{Ref6}}} & $1.74$ & $-$ & $1.72$ & $-$ & $3.28$ & $-$ \\
\multicolumn{1}{l}{(Pooling Functionals)} & 
\makecell[l]{\multirow{1}{*} {\emph{Median} \cite{weightedmedian,TAY2021108302}}} & $1.72$ & $1.97$	& $1.72$ & $2.02$ & $3.24$ & $3.73$ \\& 
\makecell[l]{\multirow{1}{*} {\emph{Middle} \cite{Ref6}}}& $1.72$ & $-$	& $1.72$	& $-$ & $3.27$ & $-$ \\& 
\makecell[l]{\multirow{1}{*} {\emph{Last} \cite{Ref6}} }	& $1.72$	& $-$ & $1.72$ & $-$	& $3.28$ & $-$ \\& 
\makecell[l]{\multirow{1}{*} {\emph{Mean}\&\emph{Std.} \cite{Ref6}}} & $1.90$ & $2.04$ & $1.88$ & $2.05$ &		$3.72$ & $3.96$ \\ 
\midrule
\makecell[l]{\multirow{2}{*} {GNN-Based} }& \makecell[l]{\multirow{1}{*}{Graph U-Net \cite{Ref20}}}& $1.78$ & $1.87$ & $1.68$ & $1.98$ & $3.19$ & $3.78$ \\
& \makecell[l]{\multirow{1}{*}{GATcosine \cite{Ref29}}}& $1.85$ & $1.76$ &	$1.73$ & $1.74$ &	$3.40$ & $3.26$ \\
\midrule
\makecell[l]{\multirow{1}{*} {GRU-Based} }& \makecell[l]{\multirow{1}{*}{GRU (RawNet2)~\cite{Ref44}}}& $2.07$ & $1.98$ & $2.05$ & $1.96$ & $3.94$ & $3.81$ \\

\midrule
\makecell[l]{\multirow{4}{*} {Attentive-Based} }

 & \makecell[l]{\multirow{1}{*}{ABP \cite{RefABP}}} & $1.76$ & $2.28$ & $1.78$ & $2.26$ & $3.30$ & $4.12$ \\

 & \makecell[l]{\multirow{1}{*}{TAP \cite{RefTAP}} } & $\bf{1.71}$ & $2.58$ & $1.72$ & $2.57$ & $3.28$ & $4.64$ \\
 & \makecell[l]{\multirow{1}{*}{SAP \cite{RefSAP}}} & $1.78$ & $1.89$ & $1.76$ & $1.72$ & $3.45$ & $3.50$ \\
& \makecell[l]{\multirow{1}{*}{ASP \cite{RefASP}}} & $1.96$	& $-$ & $1.88$ & $-$	& $3.69$ & $-$ \\
\midrule
\multicolumn{2}{l|}{\xx{ECAPA-TDNN ~\cite{Chen2022-LSS}}} & $ \xx{2.21}$ & $ \xx{1.83}$ & $ \xx{2.23}$ & $\xx{1.87}$ & $\xx{4.16}$ & $\xx{3.85}$ \\
\midrule
\multicolumn{2}{l|}{\textbf{IsoGAT (Proposed)}} 
& $1.76$ 
& $\bf{1.61}$ & $\bf{1.66}$ & $\bf{1.57}$ & $\bf{3.17}$ & $\bf{3.01}$ \\
\bottomrule

\end{tabular}
	\begin{tablenotes}
			\footnotesize
			\item 1: The EER results represented as `$-$' indicate the corresponding approaches do not obtain comparable results.
	\end{tablenotes}
\end{threeparttable}

\end{table*}

\noindent\textbf{Theorem 2:} The vertex aggregation representation of Equation (\ref{GINAGG}) is injective.

\noindent\textbf{Proof:} We first induce Lemma 1 as presented and proved in~\cite{GIN}, in order to help on the proof.

\noindent \textbf{Lemma 1:}  For a given countable multiset $\mathcal{X}$, there exists a function $g:\mathcal{X} \rightarrow \mathbb{R}^n$ such that $\psi (c,\mathcal{Z} )=(1+\epsilon )g(c)+\sum_{z\in \mathcal{Z} }{g}(z)$ is unique for each pair $(c,\mathcal{Z} )$ ($c \in \mathcal{X}$), where $\mathcal{Z}  \subset \mathcal{X}$ is a finite multiset and $\epsilon$ can be any numbers, including irrational numbers. Any function $\gamma$ over such pairs can be decomposed as $\gamma(c,\mathcal{Z} )=\phi \left( (1+\epsilon )g(c)+\sum_{z\in \mathcal{Z} }{g}(z) \right)$ for some function $\phi(\cdot)$. 

We further set $g(c)=c$ and $g(z)=z$ to an arbitrary dimension of $a_{i,i}\boldsymbol{h}_{i}^{\left( k-1 \right)}$ and $a_{i,j}\boldsymbol{h}_{j}^{(k-1)}$, respectively, where $z$'s candidate $\mathcal{Z}$ corresponds to the arbitrary dimension for the weighted hidden-state set $\{a_{i,1}\boldsymbol{h}_{1}^{(k-1)},a_{i,2}\boldsymbol{h}_{2}^{(k-1)},\ldots,a_{i,i-1}\boldsymbol{h}_{i-1}^{(k-1)},a_{i,i+1}\boldsymbol{h}_{i+1}^{(k-1)},\ldots,a_{i,N}\boldsymbol{h}_{N}^{(k-1)}\}$. 
Then, through setting $\phi(\cdot)$ to $f^{(k)}(\cdot)$, it can be proved that each element in $\boldsymbol{h}_{i}^{(k)}$ is unique, and hence, Theorem 2 can be proved when considering all the dimensions. $\blacksquare$

Finally, we present a readout phase through setting a readout function $g(\cdot)$ on the embeddings' hidden states, as 
\begin{equation}
\boldsymbol{h}^{(k)}=g\left(\mathcal{H}^{(k)} \right) =\frac{1}{2}\left( \frac{1}{N}\sum\nolimits_{i=1}^N{\boldsymbol{h}_{i}^{(k)}} +Median\left(\mathcal{H}^{(k)} \right) \right),
\end{equation}
where the state set $\mathcal{H}^{(k)}=\{\boldsymbol{h}_{1}^{(k)},\boldsymbol{h}_{2}^{(k)},\ldots,\boldsymbol{h}_{N}^{(k)}\}$ and $Median(\cdot)$ indicates the median value of the set. 

As shown in Figure~\ref{aggregation_module}, We apply the readout function to all the layers' hidden states and the weighted-sum results, obtaining the final embedding for sample $\boldsymbol{x}$ as 
\begin{equation}\label{eq9}
\boldsymbol{z}=\frac{1}{\sum\nolimits_{k=0}^K{(u^{(k)}+v^{(k)})}}\sum\nolimits_{k=0}^K\left(u^{(k)}g(\mathcal{H}^{(k)})+v^{(k)}g(\mathcal{M}^{(k)})\right),
\end{equation}
where $u^{(k)}$ and $v^{(k)}$ are the learnable weights initilized with $1$. 
$\mathcal{M}^{(k)}=\{\boldsymbol{m}_{1}^{(k)},\boldsymbol{m}_{2}^{(k)},\ldots,\boldsymbol{m}_{N}^{(k)}\}$ represents the set of the weighted sum for the $k$th layer, leading to the corresponding $k$th-layer result written as 
\begin{equation}
\boldsymbol{m}^{(k)}=g\left(\mathcal{M}^{(k)} \right) =\frac{1}{2}\left( \frac{1}{N}\sum\nolimits_{i=1}^N{\boldsymbol{m}_{i}^{(k)}} +Median\left(\mathcal{M}^{(k)} \right) \right).
\end{equation}

\section{Experimental Setups}
\label{sec-es}

\subsection{The Datasets}
\label{subsec-TD}

In order to evaluate the performance of the proposed IsoGAT for SR, \xx{we perform experiments on the VoxCeleb1\&2\footnote{https://www.robots.ox.ac.uk/$ \sim $vgg/data/voxceleb/vox1.html}\textsuperscript{,}\footnote{https://www.robots.ox.ac.uk/$ \sim $vgg/data/voxceleb/vox2.html} datasets~\cite{vox1, vox2}. }
The development set of VoxCeleb2 contains approximately $1.1$ million utterances from $5\,994$ celebrates, collected in about $146$ thousand videos from the YouTube platform, with the average speech duration of $7.2$ seconds. 
A validation set is created based on about $2\%$ of the development set, including all the speakers without overlapping in recordings.

The reported performance is measured in terms of \emph{Equal Error Rates} (EERs) considering inter-speaker and intra-speaker discrimination, evaluated on clean original (Vox1-o), extended (Vox1-e), and hard (Vox1-h) test sets from the VoxCeleb1 data, with the speaker and sample information shown in Table~\ref{dataset}. 
Note that we keep a speaker-independent setup between the development (VoxCeleb2) and the test (VoxCeleb1) sets. 
Further, the pre-trained weights used in the experiments within the wav2vec 2.0 framework have been made available on Hugging-Face\footnote{https://huggingface.co/facebook/wav2vec2-base}\label{pretrain-weight}.


\subsection{Implementation Details} 


In the graph attention module, we set $F'^{(k)}=F'^{(0)}=F$, without change on the dimensionality. Further, we set $K=1$ in the aggregation module, considering one hidden layer for each MLP with $1\,024$ hidden nodes. We set $\epsilon=0$ in the aggregation scheme, in accordance with the setting in~\cite{GIN}.
The loss function of the network is \emph{Additive Angular Margin} (AAM) softmax loss~\cite{Ref35, Ref36}, with a scale of $30$ and a margin of $0.2$. 
We also employ the cosine similarity\footnote{https://github.com/mravanelli/pytorch-kaldi} as the back-end performance evaluation tool.

 

In each experiment, we set the batch size to $48$ and set each sample’s duration to $3$ seconds sampling from the audio files, without data augmentation techniques. 
To expedite the convergence process of our model, we implement a two-stage fine-tuning strategy and first fine-tune mean-pooling model for $30$ epochs with the learning rate of $10^{-5}$ on the first feature selection approach, leveraging the pre-trained weights of wav2vec 2.0~\cite{Ref3,Ref6}. Then, we utilize the initial weights for the pre-trained mean-pooling model, undergoing $5$ training epochs with the same learning rate $10^{-5}$. 
The optimizer is set to \emph{Adaptive moment estimation} (Adam)~\cite{Ref33} with a OneCycle learning rate schedule  \cite{Ref34}.


\section{Experimental Results}
\label{sec-er}

\subsection{Experimental Comparisons}\label{secBaseline}

In the experiments, we aim to make comparison between the proposed and existing approaches when using the self-supervised pre-trained model of wav2vec 2.0, as shown in Table~\ref{results}, where `Last Layer'  and `All Layers' refer to using the last-layer and all the layers as the input representation, respectively, as in Section~\ref{subsec-RLM}. The compared approaches include classical pooling strategies (using temporal-pooling functionals), GNN-based approaches, RNN-based approaches, and attentive-based approaches. Note that the GNN-based, RNN-based, attentive-based, and proposed approaches employ adaptive strategies with learnable pooling. 

\noindent\textbf{Comparisons with Classical Pooling}

First, we aim to investigate the performance between the proposed approach and the classical pooling with fixed processing strategies for the low-level embeddings. We present the classical-pooling results corresponding to mean~\cite{Ref4, Ref6}, maximum~\cite{Ref6}, random~\cite{Ref6}, first-embedding (noted as `\emph{First}')~\cite{Ref6}, median~\cite{weightedmedian,TAY2021108302}, middle-embedding (noted as `\emph{Middle}')~\cite{Ref6}, last-embedding (noted as `\emph{Last}')~\cite{Ref6} pooling, and mean pooling with standard deviation (noted as `\emph{Mean}\&\emph{Std.}'). 

It can be drawn from the results that the proposed IsoGAT can achieve better performance (with the EER results of $1.61\%$, $1.57\%$, and $3.01\%$ for the datasets of Vox1-o, Vox1-e, and Vox1-h, respectively) compared with the classical-pooling approaches, especially for the all-layer setting. Nevertheless, for the last-layer setting, the proposed approach may not lead to dominant performance. This is probably because the classical-pooling approaches may fail to effectively describe complex data including influences from negative information.

\begin{table}[t]\small
\centering
\begin{threeparttable}
\caption{\xx{The numbers of parameters contained in the models for their corresponding approaches.}}
\label{parameters}
\begin{tabular}{l c c}
\toprule
\textbf{Models} & \makecell{\textbf{Representation }\\ \textbf{Fusion}} \\ 
\midrule
GRU (RawNet2) \cite{Ref44} & $7.9$M  \\
Classical Pooling & \multirow{2}{*}{$-$} \\
(excl. \emph{Mean}\&\emph{Std.})~\cite{Ref4,Ref6} & & \\
Graph U-Net \cite{Ref20} & $592$K  \\
Classical Pooling & \multirow{2}{*}{$-$}  \\
(\emph{Mean}\&\emph{Std.})~\cite{Ref6} & & \\

ASP \cite{RefASP} & $394$K\\
TAP  \cite{RefTAP}& $590$K \\
SAP \cite{RefSAP} & $1.2$M \\
ABP  \cite{RefABP} &$1$ K\\
\xx{ECAPA-TDNN~\cite{Chen2022-LSS}} &\xx{ 6M }\\
IsoGAT (Proposed) & $2.2$M\\
\bottomrule

\end{tabular}

	\begin{tablenotes}
			\footnotesize
			\item 1: The `$-$'s indicate that the corresponding modules do not introduce new parameters.
			
	\end{tablenotes}
\end{threeparttable}
\end{table}

\noindent\textbf{Comparisons within Learnable Pooling}

Then, we perform comparisons on the learnable-pooling approaches as in Table~\ref{results}, between the proposed IsoGAT and the other approaches on adaptive pooling. Within these approaches, the GNN-based approaches include Graph U-Net~\cite{Ref20} and GATcosine~\cite{Ref29}, while the approach using GRU (RawNet2)~\cite{Ref44} is also included in the comparisons as a typical RNN-based approach.  Afterwards, we also present attentive-based approaches including \emph{Self-Attentive Pooling} (SAP)~\cite{RefSAP}, \emph{Attentive Bilinear Pooling} (ABP)~\cite{RefABP}, \emph{Temporal Average Pooling} (TAP)~\cite{RefTAP},  \emph{Attentive Statistics Pooling} (ASP)~\cite{RefASP}. \xx{Further, we employ ECAPA-TDNN in the experiments~\cite{Chen2022-LSS}, as it plays a same role with the pooling modules with the proposed IsoGAT. Note that in order to make a fair comparison, we employ the same wav2vec 2.0 model and fine-tuning strategy as in IsoGAT for the ECAPA-TDNN.}


\xx{For the GNN-based, ECAPA-TDNN, and attentive-based approaches, it is learnt from Table~\ref{results} that the proposed approach performs better than the compared approaches when using the all-layer embeddings, which shows the effectiveness for the proposed IsoGAT using the self-supervised representation.}  This indicates that for the proposed IsoGAT may benefit from isomorphic graph attention for speaker recognition tasks, compared with the GNN-based and attentive-based pooling approaches.


In addition, we also observe that a GRU-based backend in the RawNet2 approach corresponds to lower EER results, compared with most of the approaches in the experiments. This may be due to the existence of the Transformer layers in the self-supervised pre-trained model, which provides sufficient encoding capacity for the model. In this regard, the inclusion of the GRU module may re-organize the pre-trained model's inherent description for the SR tasks.  

\xx{Furthermore, we observed that specific pooling methods (e.g., classical pooling, graph U-Net, and attention-based pooling methods), exhibit better performance when utilizing the last layer compared with the all-layer setup. This trend deviates from conventional expectations, with the hypothesis that this phenomenon is closely related to our fine-tuning strategy. Consequently, these methods may gain an advantage by focusing on the last layer, as it is richly endowed with speaker-relevant information.}

\begin{table}[t]\small
 \centering

 \caption{The EER ($\%$) results of IsoGAT approaches with and without the MLP layer, noted as `w/ MLP' and `w/o MLP', respectively, on the three test sets. } \label{Ablationinjective}
 \begin{tabular}{l c c c}
 \toprule
 \multicolumn{1}{c}{\textbf{Approaches}}  &  \multicolumn{1}{c}{\textbf{Vox-o}} & \multicolumn{1}{c}{\textbf{Vox-e}}&\multicolumn{1} {c}{\textbf{Vox-h}}\\
\midrule
\multirow{1}{*}{IsoGAT (w/o MLP) } & $1.64$ & $1.62$ & $3.11$ \\
\multirow{1}{*}{IsoGAT (Proposed; w/ MLP)} & $1.61$ & $1.57$ & $3.01$ \\
 \bottomrule
 \end{tabular}

 \end{table}

\begin{table}[t]\small
\centering
\caption{The EER ($\%$) results for the IsoGAT when using different numbers of layers on three test sets.}
\label{Abaltionlayers}
\begin{tabular}{l ccc}
\toprule
\textbf{\# Layers ($K$)} & \textbf{Vox-o} & \textbf{Vox-e} & \textbf{Vox-h} \\ \midrule
$K=1$ (Proposed) & $\bf{1.61}$ & $\bf{1.57}$ & $\bf{3.01}$ \\
$K=2$& $1.63$ & $1.62$ & $3.04$ \\
$K=3$ & $1.69$	& $1.67$ & $3.15$ \\

\bottomrule
\end{tabular}
\end{table}

\noindent\textbf{Discussion} 

\begin{figure}[t]
\centering
\includegraphics[width=3.4in]{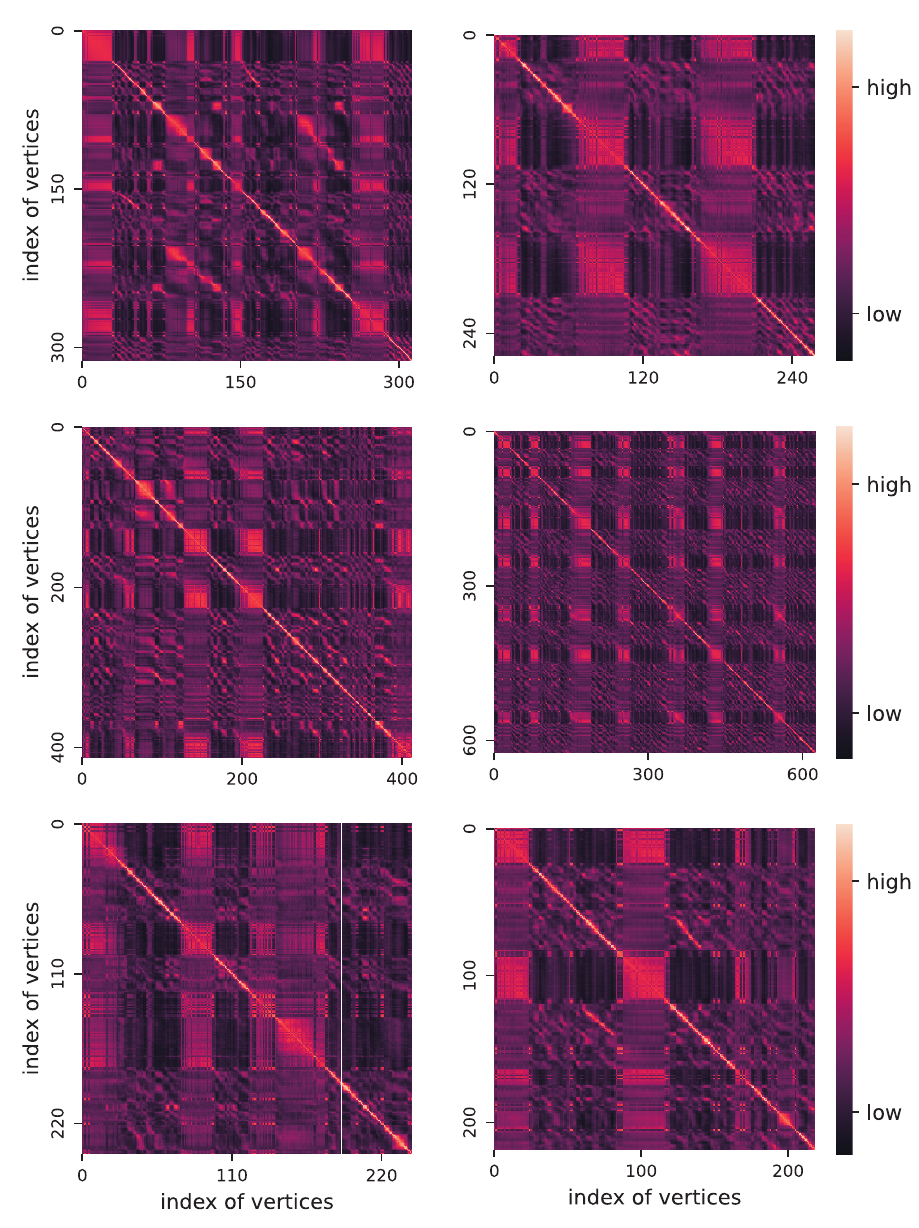}
\caption{The visualized adjacency matrices corresponding to six utterances, respectively, from different speakers, where the brighter pixels represent larger values of the adjacency weights.}
\label{weightmatirx}
\end{figure}

Since the proposed IsoGAT implicitly builds adjacency matrices in training its models, we further visualize the adjacency matrices in Figure~\ref{weightmatirx} with each of its subfigures corresponding to a speaker's utterance, showing the weight elements of the adjacency matrices. Note that the vertical and horizontal axes in Figure~\ref{weightmatirx} indicate the index of vertices with respect to the low-level embeddings within an utterance. 
As can be seen from the figure, larger weights frequently appear in some neighboring embeddings within an utterance, which indicates that the speaker-identity information in the utterance may obey a sparse distribution. 
Further, the large adjacency weights' `block-wise' distribution also implies that, the proposed approach tends to model these embedding into temporally discrete forms, which is different from the classical-pooling strategies and the sequential description for a GRU-based approach.

 Then, we present the parametric sizes of different phases for the proposed and compared approaches in the experiments, as shown in Table~\ref{parameters}. Note that the numbers of the parameters are different between the `\emph{Mean}\&\emph{Std.}' and the other setups (noted as `excl. \emph{Mean}\&\emph{Std.}') in the classical-pooling approaches. The comparison indicates that the proposed IsoGAT can achieve better speaker-recognition performance with the amount of the parameters in the same order of magnitude.

\subsection{Ablation Study}

\noindent\textbf{Influence of the Aggregation Module}

First, we shed light on the influence of the MLP block contained in each layer for the aggregation module. 
As presented in Table~\ref{Ablationinjective}, we perform experiments on the three evaluation sets and make comparison between the IsoGAT with and without the MLP block (noted as `w/ MLP' and `w/o MLP', respectively) using the EER indicators. 
It is learnt from the table that the proposed IsoGAT slightly outperforms its without-MLP setup, benefits from the inclusion of the MLP block for SR tasks.  



\begin{table}[t]\small
 \centering
\begin{threeparttable}
 \caption{The test-set EER ($\%$) results of different pooling approaches tuning on the original pre-trained weights of wav2vec 2.0.} \label{Ablationweights}
 \begin{tabular}{l  c c c}
 \toprule
 \multicolumn{1}{c}{\textbf{Approaches}} & \multicolumn{1}{c}{\textbf{Vox-o}} & \multicolumn{1}{c}{\textbf{Vox-e}}&\multicolumn{1} {c}{\textbf{Vox-h}}\\\hline
Last Layer: & & &\\\hline
\emph{Mean} Pooling (w/o M. Tuning)~\cite{Ref4, Ref6} & $3.24$ & $3.50$ & $6.72$\\
Graph U-Net (w/o M. Tuning)~\cite{Ref20} & 
  $2.55$ & $2.96$ & $\bf{5.50}$  \\
IsoGAT (w/o M. Tuning) & $\bf{2.52}$ & $\bf{2.82}$ & $5.56$  \\\hline

All Layers: & & &\\\hline
\emph{Mean} Pooling (w/o M. Tuning)~\cite{Ref4, Ref6} & $2.36$ & $2.71$ & $5.22$ \\
Graph U-Net (w/o M. Tuning)~\cite{Ref20} & $2.41$ & $2.62$ & $5.12$  \\
IsoGAT (w/o M. Tuning) & $\bf{2.11}$ & $\bf{2.35}$ & $\bf{5.00}$  \\
 \bottomrule
 \end{tabular}
\end{threeparttable}
 \end{table}

Then, we aim to investigate the influence of the total number of the layers ($K$) in the aggregation module, since we choose $K=1$ in the proposed IsoGAT. To this end, we show the EER performance in Table~\ref{Abaltionlayers} for the cases of $K$ set to $1$, $2$, and $3$, respectively, when considering the setup of `All Layers' as in Table~\ref{results}. 
Through observing the results, it can be drawn that increasing the number of the layers failes to definitely result in improvement on SR performance, in view of the best EER results for $K=1$. 
This is possibly due to the fact that we only consider to construct fully-connected graphs in the proposed approach, which implies that learning on first-order neighbors is sufficient for obtaining optimal models.

\noindent\textbf{Influence of the Pre-Trained Weights}

Although our proposed IsoGAT achieves better performance compared with the other pooling approaches, it still remains arguable on whether the improvement of IsoGAT results from the graph-learning strategies, since the proposed IsoGAT jointly perform optimization for the 30-epoch mean pooling pre-trained weights. 
Hence, we present Table~\ref{Ablationweights} containing the EER results for the approaches of classical mean pooling, Graph U-Net, and IsoGAT without tuning on the mean pooling pre-trained weights (noted as `w/o M. Tuning') in the self-supervised wav2vec 2.0 model, when considering the last-layer and all-layer setups. Note that we directly employ the pre-trained weights released in Hugging-Face (see Section~\ref{subsec-TD}). 

It can be concluded from the EER results in Table~\ref{Ablationweights} that when direcly employing the pre-trained weights, the proposed IsoGAT also outperforms the existing approaches in most cases. This indicates that the better performance achieved by IsoGAT can attributes to the inclusion of the graph-learning based components, i.e., the graph attention with aggregation. 


\section{Conclusion}
\label{sec-con}

In this paper, we propose a speaker recognition approach using \emph{Isomorphic Graph ATtention network} (IsoGAT) on self-supervised representation, aim at solving the problems of fixed and non-injective pooling within existing approaches. The proposed approach contains the representation learning, graph attention, and aggregation modules, in order to learn optimal representation for spoken signals adapting to speakers' identities. Then, experimental results evaluated on different datasets indicate that the proposed approach achieves better performance for speaker recognition, compared with state-of-the-art pooling approaches based on self-supervised representation. 

Our future works may focus on two aspects as follows. First, we expect to investigate temporally local fusion on the self-supervised representation for speaker recognition tasks, in order to obtain sufficient multi-scale information. Second, knowledge transfer from speaker recognition to other tasks may be worth investigating, through regarding speaker recognition as an upstream task for other downstream processing.

\bibliographystyle{elsarticle-num} 
\bibliography{Ref1.bib}

\end{document}